# Coherent Radiation in Insertion Devices-II

E.G.Bessonov[*)], A.A. Mikhailichenko[**)]

[*)]Lebedev FIAN, Moscow, 119991
[**)]Cornell University, CLASSE, Ithaca, New York, 14853

*Abstract.* We represent results of calculations of coherent synchrotron radiation (CSR) of the relativistic bunch in an undulator with a vacuum chamber of arbitrary cross section with a new algorithm. This algorithm associated with direct calculations of electric field rather than the vector potential. CSR normalized to the incoherent one and compared with analytical calculations for a free space.

## INTRODUCTION

In our previous publication [1] we introduced the method of calculation of coherent radiation in an undulator. The method described in [1] dealt with calculations of vector and scalar potentials with appropriate boundary condition on the surface of vacuum chamber having arbitrary cross section. Later we changed algorithm slightly, so the scalar potential calculated from the Lorentz condition rather than from the wave equation for a scalar potential.

In the current publication we describe some details of newest algorithm for numerical calculation of power radiated coherently. In these new calculations of power we use the direct equations for the electric field, rather than equations for the vector (and scalar) potential. This delivers faster algorithm and much smoother solutions. With the help of this new algorithm we investigated CSR in an undulator suggested for ERL (see details of this undulator in [11]).

Let we briefly remind the results obtained in [1] first.

## CR IN A FREE SPACE

The particle in an undulator performs transverse oscillations with a spatial period of undulator field, which is $\lambda_u = 2\pi \lambdabar_u$. While oscillating in the undulator field, the particle radiates on harmonics [2]-[4]

$$\omega_n = \frac{n\Omega}{1 - \beta_\parallel \cos\vartheta}, \qquad (1)$$

where $n=1,2,3\ldots$, numerates the harmonics of frequency $\Omega = \beta_\parallel c / \lambdabar_u$, $\overline{\beta} = \overline{v}/c \cong \beta_\parallel$, $\overline{v}$ is a particle's average longitudinal velocity in the undulator. For a helical undulator one revolution around the axis corresponds to a single oscillation in the longitudinal direction also, so the harmonics numerator is the same for the transverse and longitudinal oscillations. Formula (1) corresponds to the Doppler-shifted frequency $\Omega$. Accordingly, the wavelength of radiation is

$$\lambdabar_n = \frac{\lambdabar_u \cdot (1 - \beta_\parallel \cos\vartheta)}{n\overline{\beta}} \qquad (2)$$

The energy distribution of undulator radiation (UR) emitted by a single particle in an undulator of length $L_u = M\lambda_u$ ($M$ –is the number of periods), during the time duration $\Delta t = 2\pi M/\Omega$ is defined by the expression [2]-[5] (see Appendix)

$$\frac{d\varepsilon_n}{do} = \frac{dI_n}{do}\frac{2\pi M}{\Omega} = \frac{e^2\omega_n^3 M}{cn\Omega^2}\left[\beta_\perp^2 J_n'^2\left(\frac{n\beta_\perp \sin\vartheta}{1-\beta_\parallel \cos\vartheta}\right) + \frac{(\cos\vartheta - \beta_\parallel)^2}{\sin^2\vartheta} J_n^2\left(\frac{n\beta_\perp \sin\vartheta}{1-\beta_\parallel \cos\vartheta}\right)\right] \quad (3)$$

where $\beta_\perp = v_\perp/c = K/\gamma$, $v_\perp$ is the transverse velocity. In the dipole approximation, $K<1$, $\beta_\parallel = \sqrt{\beta^2 - \beta_\perp^2} \cong \beta$ and in the ultra-relativistic case $\gamma \gg 1$, mainly the first harmonic radiated, $n=1$.

$$\frac{d\varepsilon_1}{do} = \frac{e^2\Omega M}{c} \cdot F(\vartheta) \cong \frac{e^2\Omega M}{c} \cdot \frac{\beta_\perp^2}{8\pi(1-\beta_\parallel \cos\vartheta)^3}\left[1 + \frac{(\cos\vartheta - \beta_\parallel)^2}{(1-\beta_\parallel \cos\vartheta)^2}\right] \quad (4)$$

where the function $F(\vartheta)$ introduced accordingly. Indeed, for total radiation at all harmonics, expression (3) should be summarized over all indices $n$ [12], [5]

$$\frac{dI}{do} = \sum_{n=1}^{\infty}\frac{dI_n}{do} = \frac{e^2\Omega^2\beta^2}{32c} \cdot \left[\frac{4+3\beta^2\sin^2\vartheta}{(1-\beta^2\sin^2\vartheta)^{5/2}} + \frac{\cos^2\vartheta \cdot (4+\beta^2\sin^2\vartheta)}{(1-\beta^2\sin^2\vartheta)^{7/2}}\right] \quad (5)$$

The energy, radiated at first harmonic by a bunch with population $N_b$ within angles $\{\pi, \vartheta_m\}$ can be estimated by integrating (3) or (4) over the solid angle

$$\Delta\varepsilon_1 = N_e \int_{\vartheta_m}^{\pi} \frac{d\varepsilon_1(\vartheta)}{do} do = 2\pi N_e \int_{\vartheta_m}^{\pi} \frac{d\varepsilon_1(\vartheta)}{do} \sin\vartheta d\vartheta \quad (6)$$

In case $\vartheta_m = 0$, formula (5) gives the total radiated energy at all harmonics in dipole approximation

$$\Delta\varepsilon_{tot} = N_e \tfrac{2}{3} r_e^2 \overline{B^2} \gamma^2 M\lambda_u = \tfrac{8\pi^2}{3} e^2 N_e M K^2 \gamma^2 / \lambda_u \quad (7)$$

For the energy radiated coherently at first harmonic, one should suggest $\vartheta_m = \vartheta_{coh}$, where the angle of coherence $\vartheta_{coh}$ is defined from (2), [8]. By suggesting that the wavelength of radiation coincides with the bunch sigma $\sigma_b$,

$$\sigma_b \cong \lambdabar_{coh} = \lambdabar_u \cdot (1-\beta_\parallel \cos\vartheta_{coh})/\beta_\parallel \quad, \quad (8)$$

one can calculate the angle which delivers corresponding wavelength as the following

$$\vartheta_m = \vartheta_{coh} \cong \arccos\left(1 - \frac{\sigma_b}{\lambdabar_u}\right) \quad (9)$$

Generally, the energy-loss ratio is

$$\frac{\Delta \varepsilon_{1coh}}{\Delta \varepsilon_{1tot}} \cong N_e \int_{\theta coh}^{\pi} F(\vartheta) do \Big/ \int_0^{\pi} F(\vartheta) do . \quad (10)$$

This approach is close to the one used in [6]. In this former publication the transition conditions between the coherent and non-coherent radiation regimes investigated also. The authors came to conclusion that the wavelength, which defines a transition between the coherent and noncoheret radiation is

$$2\pi \hat{\lambda}_{coh} \cong \frac{4\pi}{\beta_{\parallel} \sqrt{2\ln 2}} \cdot \sigma_b \cong 10.67 \cdot \sigma_b \text{ , i.e. } \hat{\lambda}_{coh} \cong 1.7 \sigma_b \quad (11)$$

(not a function of the beam energy); the *width* of transition region is $\Delta \hat{\lambda}_{coh} \cong 0.26 \hat{\lambda}_{coh}$, i.e. pretty narrow.

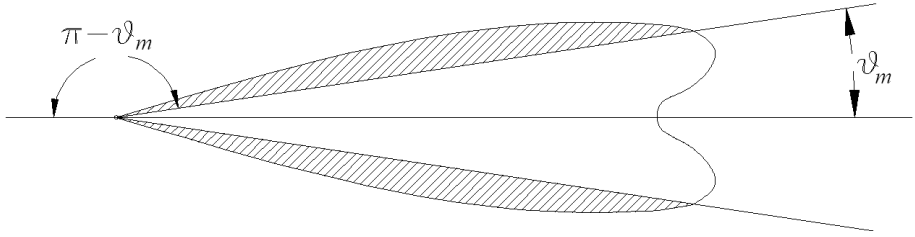

Figure 1. To the integration for comparison of coherent/incoherent ratio with formula (6). Hatched area linked to the coherent part of radiation.

Calculation with formula (10) could be done with MATHEMATICA. Some results are represented in Table 1 below. We would like to underline here that for a single electron the ratio of the total energy radiated in a forward and in a backward direction strictly is $\sim \gamma^4$ in full agreement with formula (73,11) from [7].

*Second and higher harmonics*

For the harmonic with index *n*, the wavelength is *n* times shorter, so the angle of coherence defined as

$$\vartheta_{m,n} = \vartheta_{coh,n} \cong \arccos\left(1 - \frac{n\sigma_b}{\hat{\lambda}_u}\right) \quad (12)$$

is smaller. The cut-off angle reduction becomes compensated by the angular distribution dependence however. This dependence has a maximum, shifted towards the larger angles. For the second harmonic

$$\frac{d\varepsilon_2}{do} = \frac{4e^2 \Omega M}{c(1 - \beta_{\parallel} \cos\vartheta)^3} \left[ \beta_{\perp}^2 J_2'^2 \left(\frac{2\beta_{\perp} Sin\vartheta}{1 - \beta_{\parallel} Cos\vartheta}\right) + \frac{(Cos\vartheta - \beta_{\parallel})^2}{Sin^2\vartheta} J_2^2 \left(\frac{2\beta_{\perp} Sin\vartheta}{1 - \beta_{\parallel} Cos\vartheta}\right) \right] \quad (13)$$

Expanding derivative according $J_n'(x) = \frac{1}{2}(J_{n-1}(x) - J_{n+1}(x))$ [13], equation (13) could be represented as

$$\frac{d\varepsilon_2}{do} = \frac{4e^2 \Omega M}{c(1-\beta_\parallel \cos\vartheta)^3} \left[ \frac{\beta_\perp^2}{4} \left[ J_1\left(\frac{2\beta_\perp \sin\vartheta}{1-\beta_\parallel \cos\vartheta}\right) - J_3\left(\frac{2\beta_\perp \sin\vartheta}{1-\beta_\parallel \cos\vartheta}\right) \right]^2 + \frac{(\cos\vartheta - \beta_\parallel)^2}{\sin^2\vartheta} J_2^2\left(\frac{2\beta_\perp \sin\vartheta}{1-\beta_\parallel \cos\vartheta}\right) \right], (14)$$

which could be easily expanded with MATHEMATICA. Calculation show, that the ratio

$$\frac{\Delta\varepsilon_{2coh}}{\Delta\varepsilon_{2tot}} = \frac{N_e \int_{\vartheta_{2m}}^{\pi} \frac{d\varepsilon_2(\vartheta)}{do} do}{\int_0^{\pi} \frac{d\varepsilon_2(\vartheta)}{do} \sin\vartheta d\vartheta} \quad (15)$$

remains $\sim \leq 2 \cdot 10^{-4}$ (see Table 1). So the coherent radiation at second (and higher harmonics) is extremely low and could be neglected for our purposes.

It is interesting, that for the beam duration of approximately 0.1*ps*, *Q*=100*pC*, the ratio of coherent to incoherent energy losses comes to ~0.6 for an undulator with *K*=1.5, $\lambda_u = 2cm$ at 5 *GeV*; for *K*=3 the coherent loss is 6.5 times bigger, than the incoherent one. For 1 *nQ* this ratio comes to be 65 for the last set of parameters. For *atto*second bunches radiation is purely coherent and it has the wavelengths on the order of the bunch length, which makes usage of such bunches problematic for radiation with desired properties. All these were preliminary estimations. For a more correct evaluation of the effect in a chamber of arbitrary cross section we used the FlexPDE solver [14] (see below).

## CR IN A VACUUM CHAMBER

In the presence of a vacuum chamber with rectangular cross section *a*x*a*, the wavelength of radiation $\lambda_{wg}$ becomes longer, than the wavelength $\lambda_0$, measured in a free space, $\lambda_{wg} = \lambda_0 / \sqrt{1-(\lambda_0/\lambda_{cr})^2}$, where $\lambda_{cr} \cong 2a$ (for a round chamber and H$_{01}$ wave $\lambda_{cr} \cong 1.64a$ and so on). Radiation can propagate within an angle that is not smaller than the one, defined by the critical wavelength in a waveguide $\theta_{cr} \geq arcCos(\lambda_0/\lambda_{wg})$ [9], [10]. On the other hand, according to (2), the angle corresponding to the coherence length $\lambda_{coh} \cong \sigma_b$ is $\theta_{coh} \geq arcCos\ (1 - \overline{\beta}\sigma_b / \lambdabar_u)$. This should be larger, than $\theta_{cr}$.

In a chamber with arbitrary cross section, the allowed wavelengths have rather complicated structure, so analytical solution can only be obtained for the simplest cases, such as vacuum chambers of either round or rectangular cross section. We therefore developed a code able to evaluate radiation in a conducting chamber of arbitrary cross section.

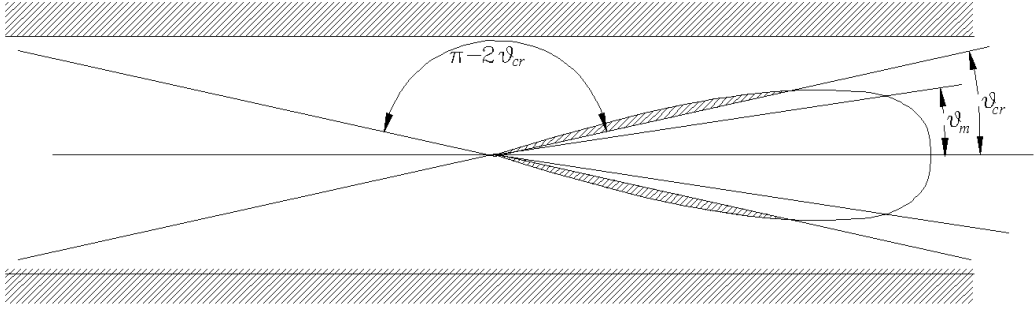

Figure 2. Radiation (including the coherent one) in a vacuum chamber can propagate within the angle $\pi - \theta_{cr} \geq \vartheta \geq \theta_{cr} = arcCos(\lambda_0 / \lambda_{wg})$.

## NUMERICAL MODEL

The current density similar to [1] was represented as the following

$$\vec{j} = \rho \cdot \vec{v} \cong \rho(x,y,z,t) \cdot (v_x, v_y, v_z), \tag{16}$$

where the charge density

$$\rho(\vec{r},t) = \frac{Q_b}{\pi^{3/2} \sigma_x \sigma_y \sigma_b} \exp\left(-\frac{(z+z_0-\beta ct)^2}{\sigma_z^2}\right) \exp\left(-\frac{(x-x_0)^2}{\sigma_x^2}\right) \exp\left(-\frac{y^2}{\sigma_y^2}\right), \tag{17}$$

and

$$v_x = cK/\gamma \cdot Cos(z/\lambdabar_u), \qquad v_y = cK/\gamma \cdot Sin(z/\lambdabar_u) \tag{17a}$$

or $v_y = 0$ for a planar undulator, $z = v_z t$; $v_z \cong \beta_\parallel c = c\sqrt{\beta^2 - \beta_\perp^2} = c\sqrt{\beta^2 - K^2/\gamma^2}$, $\beta = \sqrt{1 - 1/\gamma^2}$, $\sigma_x, \sigma_y$ are the transverse sigmas in $x$ and $y$ direction $\sigma_z$ is a longitudinal sigma. The current distribution (11), (12) allows modeling the coherent part of radiation only. The other type of distribution used is even power of *cos*. In the lowest order it is the following

$$\rho(\vec{r},t) = \begin{cases} 0, \text{ if } \dfrac{z}{\sigma_z} < \dfrac{\beta ct + z_0}{\sigma_z} - \dfrac{\pi}{2} \\ \dfrac{2Q_b}{\pi^2 \sigma_x \sigma_y \sigma_z} \cos^2\left(-\dfrac{z - z_0 - \beta ct}{\sigma_z}\right) \times \exp\left(-\dfrac{(x-x_0)^2}{\sigma_x^2} - \dfrac{y^2}{\sigma_y^2}\right) \\ 0, \text{ if } \dfrac{z}{\sigma_z} > \dfrac{\beta ct + z_0}{\sigma_z} + \dfrac{\pi}{2} \end{cases} \tag{18}$$

The graphs for (17) and (18) together are represented in Fig.3.

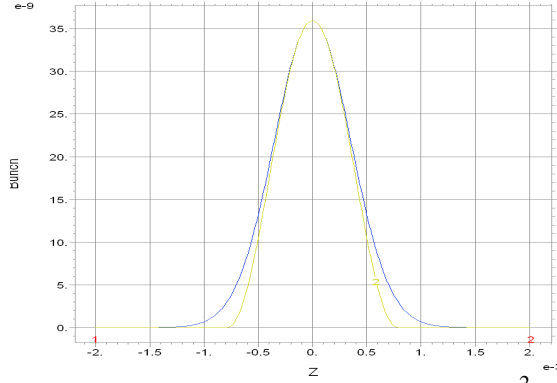

Figure 3. Gaussian distribution (17), the blue line, vs $\cos^2$ (18), the green line.

We solved the equations for the vector potential

$$\Delta \vec{A} - \frac{1}{c^2}\ddot{\vec{A}} - \delta \cdot \dot{\vec{A}} = -\mu_0 \vec{j}, \qquad (19)$$

where $\mu_0$ stands for magnetic permeability of vacuum, with the boundary conditions as the following: tangential component of the vector potential was chosen to be zero on the boundary, so its normal derivative:

$$\vec{A}_\tau = 0, \quad \frac{\partial \vec{A}}{\partial n} = 0, \qquad (20)$$

where $\partial/\partial n$ stands for the derivative, taken along the normal direction at this point.

Instead of solving equation for a scalar potential

$$\Delta U - \frac{1}{c^2}\ddot{U} = -\frac{1}{\varepsilon_0}\rho, \qquad (21)$$

where $\varepsilon_0$ stands for electric permittivity of vacuum, we are defining the scalar potential from the Lorentz condition as equations (4) and (6) are linked by this condition as the following

$$\frac{\partial U}{\partial t} = -c^2 \mathrm{div}\,\vec{A} \qquad (22)$$

with the boundary condition $U|_\Sigma = 0$. As this is the first order equations it is calculated much faster and more smother. We would like to mention here, that the scheme, where one calculates the scalar potential with (21), further cannot calculate the vector potential with (22) as the scalar equation is not enough to restore a vector.

The mesh propagates in accordance with the bunch position. Initial mesh density was chosen either by the implemented code mesh creator or by setting the initial mesh density as

$$\sim 4 \cdot 10^3 \cdot exp\,(-4 \cdot 10^5 \cdot (x^2 + y^2));$$

see Fig 4. In addition, while the bunch propagates through the undulator the mesh becomes denser, if required by accuracy of calculations. So the time required for the next time step increased accordingly; this increase in case of exponential mesh density is minimal however.

The term with decrement, $\delta \cdot \dot{\vec{A}}$, $\delta = 1/c^2\tau$, with $\tau \sim 1 ns$, is introduced to describe the losses in the walls. Although the losses occur at the surface, the volume losses introduced by this way are self consistent. In our model of vacuum chamber, the end section, at ~5% along the z-distance at the entrance, has increased losses, ~$100\delta$ to avoid reflections from the entrance boundary (this section marked by the blue color in Fig. 4).

Magnetic field defined from vector potential in the first sets of runs as

$$\vec{B} = curl\vec{A}, \tag{23}$$

and the electric field defined as

$$\vec{E} = -\frac{\partial \vec{A}}{\partial t} - gradU. \tag{24}$$

Energy lost by the bunch evaluated by taking the integral

$$\Delta \varepsilon = \int dt \int (\vec{j} \cdot \vec{E}) dV = -\int dt \int \vec{j} \cdot (\frac{\partial \vec{A}}{\partial t} + gradU) dV, \tag{25}$$

Other integral calculated was taken for the vector

$$\vec{E}_\tau = \{E_x, E_y, 0\} \tag{26}$$

where the only transverse components of $\vec{E}$ are taken into account; this corresponds to pure radiation field.

$$\Delta \varepsilon_\tau = \int dt \int (\vec{j} \cdot \vec{E}_\tau) dV \tag{27}$$

*Another type of modeling*–with an electric field–was introduced finally. Namely, if the operator $\partial/\partial t$ applied to the equation (15) and operator *grad* applied to the equation (17), then after summation of these equations one can get

$$\Delta(\frac{\partial \vec{A}}{\partial t} + gradU) - \frac{1}{c^2}\frac{\partial^2}{\partial t^2}(\frac{\partial}{\partial t}\vec{A} + gradU) - \delta \cdot \frac{\partial}{\partial t}(\frac{\partial \vec{A}}{\partial t} + gradU) = -\mu_0 \frac{\partial \vec{j}}{\partial t} - \frac{1}{\varepsilon_0} grad(\rho)$$

or simply

$$\Delta \vec{E} - \frac{1}{c^2}\ddot{\vec{E}} - \delta \cdot \dot{\vec{E}} = \frac{1}{\varepsilon_0}\left(grad(\rho) + \mu_0 \varepsilon_0 \frac{\partial \vec{j}}{\partial t}\right) \tag{28}$$

with the boundary conditions $\vec{E}_\tau = 0$, $\frac{\partial \vec{E}}{\partial n} = 0$. As we are interesting in integrals (10) and (12) where the electric field appears only, the equations (24) request reduced computer power, as the number of scalar equations is three instead of six (if Lorenz condition used). In addition, as the electric field calculated from the vector potential by taking a time derivative, this procedure is extremely sensitive to the numeric algorithm for the time derivative itself, so even for a small time step, the solution with a vector potential looks fuzzier.

As the charge density $\rho$ and current density $\vec{j}$ has dependence of longitudinal coordinate and time through combination $\xi = z - v_z t$, like

$$\rho = \rho(x, y, z, t) = \rho(x, y, z - v_z t), \quad \vec{j} = \rho(x, y, z - v_z t) \cdot \vec{v}, \tag{29}$$

one can see that the longitudinal component of electric field source (a term at the right side of equation (13)) defined as

$$\frac{1}{\varepsilon_0}\frac{\partial \rho}{\partial \xi} - \mu_0 v_z^2 \frac{\partial \rho}{\partial \xi} = \frac{1}{\varepsilon_0}\frac{\partial \rho}{\partial \xi}\cdot\left(1 - \frac{v_z^2}{c^2}\right) \cong \frac{1}{\varepsilon_0}\frac{\partial \rho}{\partial \xi}\cdot\frac{1}{\gamma^2} \ , \qquad (30)$$

while the transverse component of the source can be identified as

$$\frac{1}{\varepsilon_0}\frac{\partial \rho}{\partial \xi} - \mu_0 \frac{v_z^2 K}{\gamma}\frac{\partial \rho}{\partial \xi} \cong -\frac{1}{\varepsilon_0}\frac{\partial \rho}{\partial \xi}\cdot\left(1 - \frac{v_z^2}{c^2}\frac{K}{\gamma}\right), \qquad (31)$$

which is substantially larger.

Finally, the procedure of calculation of coherent losses looks as the following. Equation (24) solved for time scale corresponding few undulator periods. While the transition process is finished, what could be seen from the picture of the field distribution (Fig. 5-4), then for the following two time moments integrals (21) and (23) are evaluated and the difference of the losses is calculated. Then the difference in losses calculated for the incoherent losses with formula (7), and compared with loss difference obtained earlier with formulas (21) and (23); the ratio represents the coherent/incoherent losses ratio, which could be compared with formula (9) for this ratio calculated in a vacuum.

One interesting phenomenon was identified here. Namely, establishment of electric field, while the bunch enters the vacuum chamber is going with reflection from the wall, so the sign of electric field across the chamber behind the bunch periodically changes its polarity.

The ratio of losses for a chamber represented in Table 1 while this ratio calculated with analytical formula (9), so attenuation by vacuum chamber found to be moderate.

Chambers with different shapes and dimensions were investigated, including the simplest one with rectangular cross section ("Small chamber" in Table 1). The wavelength of the undulator field $\lambda_u$ was varied from 1 to 3 *cm*. The bunch length was varied also (see Table 1).

Design of vacuum chamber for wigglers is sometimes problematic, as it contains so-called cleaning electrodes, what makes this chamber *multiply* connected [10]. As the backward radiation is 100% coherent and goes at central wavelength equal to the doubled period of wiggler, one may expect strong excitation of undesirable fields there.

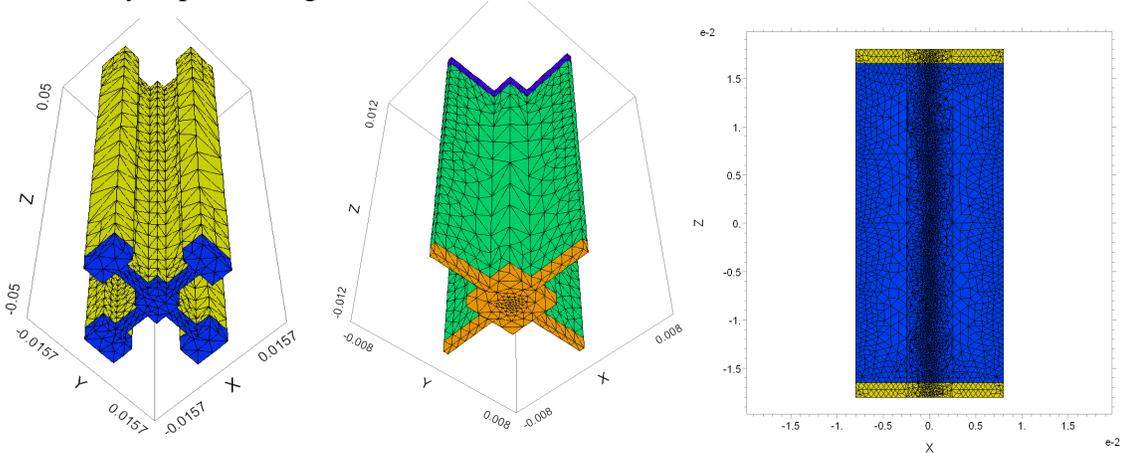

Figure 4. The chambers profiles used in a modeling, at the left, center. At the right: A transparent view through the model with exponential density of mesh.

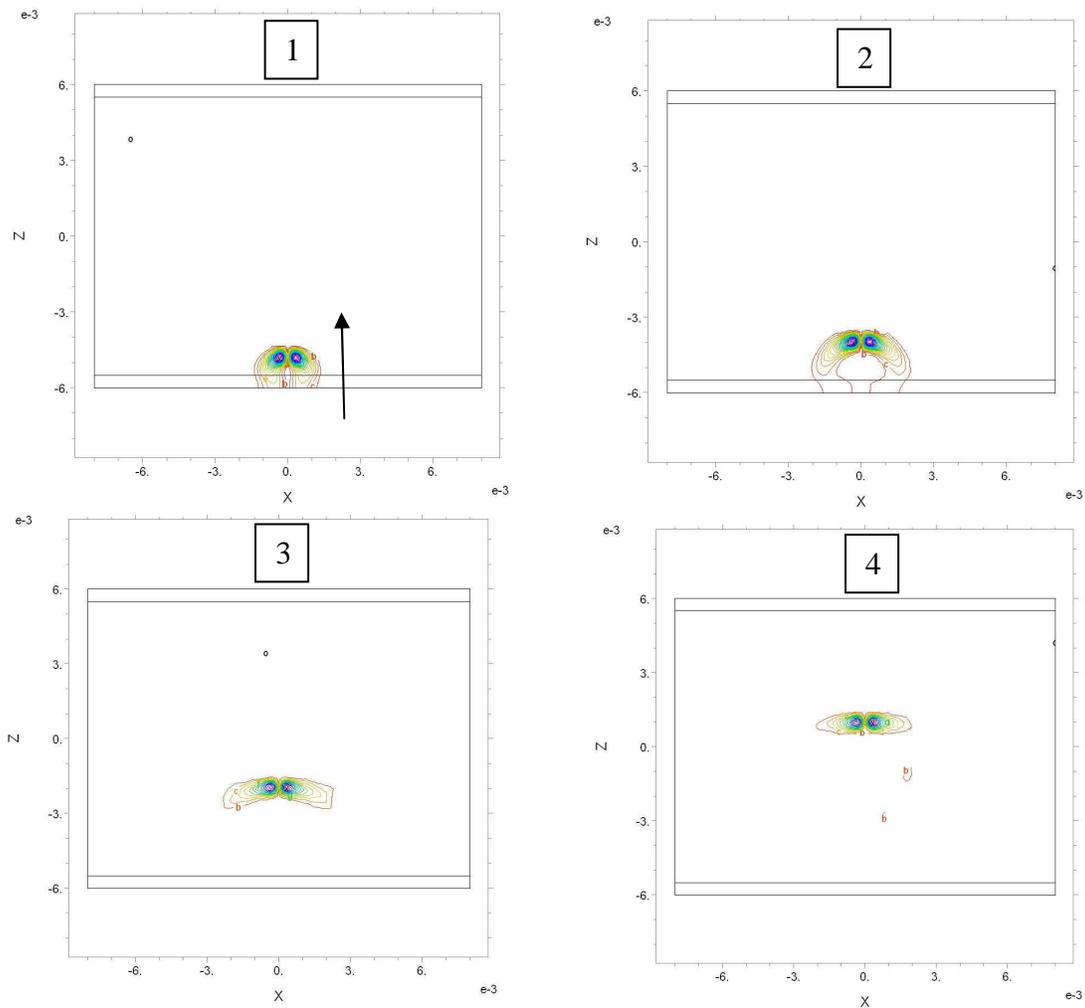

Figure 5. The electric field modulus represented for four positions of bunch (numbered). Bunch appears at the bottom of the picture at *x*=0. Longitudinal distance *z* measured in *meters*.

In Figure 5 the bunch sigma is $\sigma_z = 0.3mm$, the period of the undulator is $\lambda_u = 12mm$. So the transition ends after passage ~7 *mm*.

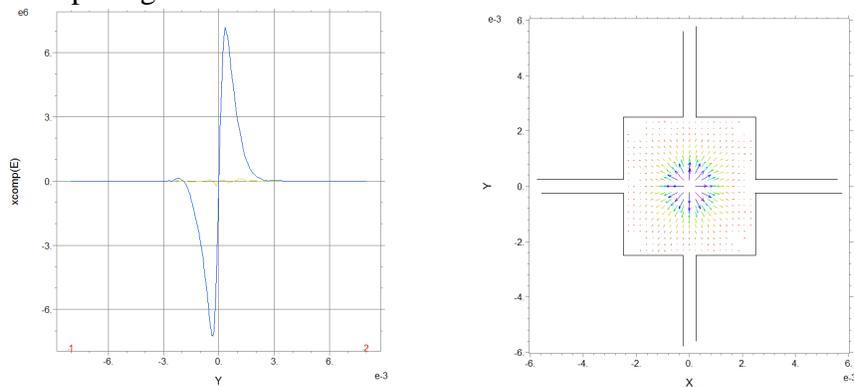

Figure 6. The electric field elevation across the bunch through its center. At the right-the electric vector field at the same cross section; (single frame from a movie).

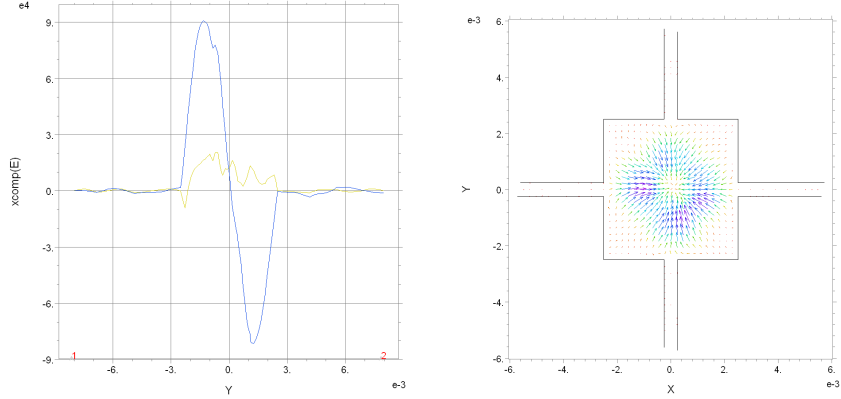

Figure 7. The electric field elevation across the bunch through $4\sigma_z$ behind its center(single frame from a movie). At the right-the electric vector field at the same cross section; compare with Fig. 6. The change of polarity of electric field is clearly seen here.

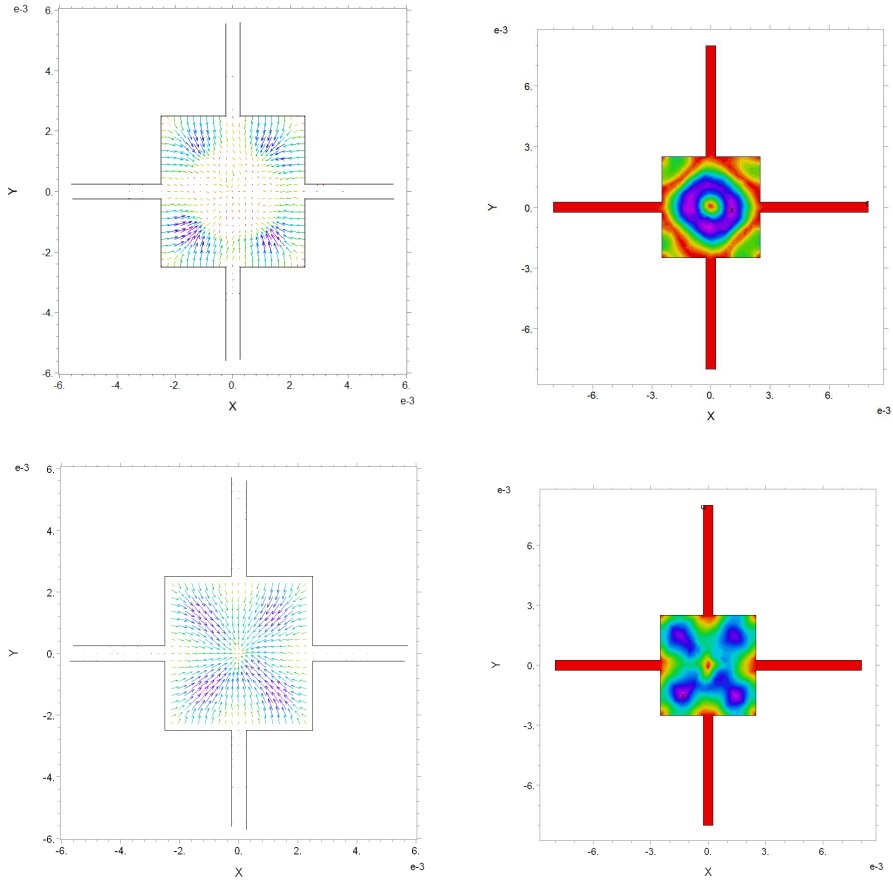

Figure8. Electric field distribution at the plane which located $4\sigma_z$ behind the center of the bunch. Vector field distribution at the left, distribution of field modulus at the right.

The field elevation across the bunch is shown in Fig.6. This pretty much as it shold be for the static cylindrical case. The bunch is moving in a chamber from Fig.4, from the viewer behind the plane of figure. Dimensions of the central part are 2.5x2.5 $mm^2$. One can see that the electric field strength is ~ 6 *MV/m* for particular parameters of the bunch.

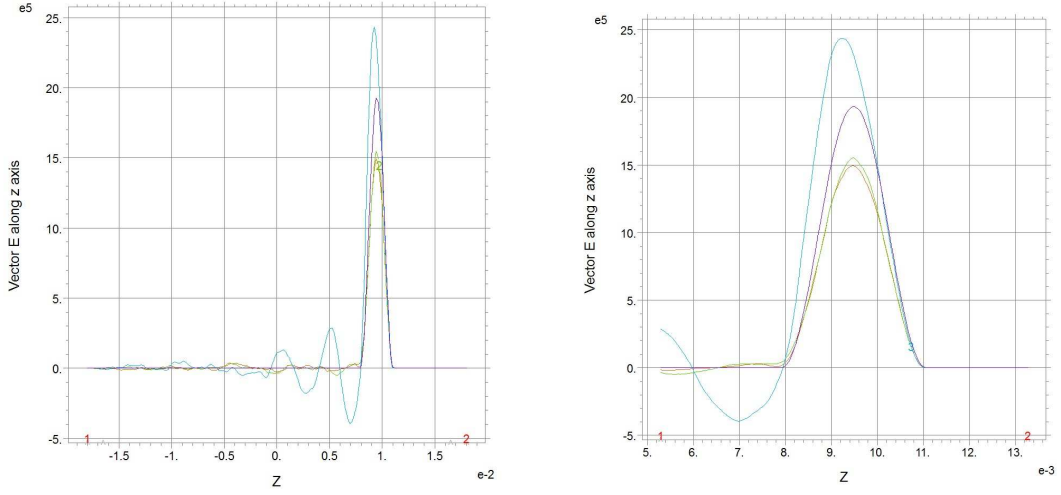

Figure 9. Components of the electric along *z*. At the right the central part of the figure zoomed at $\pm 4\sigma_z$, $\sigma_z = 1mm$. The bunch is moving from the left to the right side (frames from a movie).

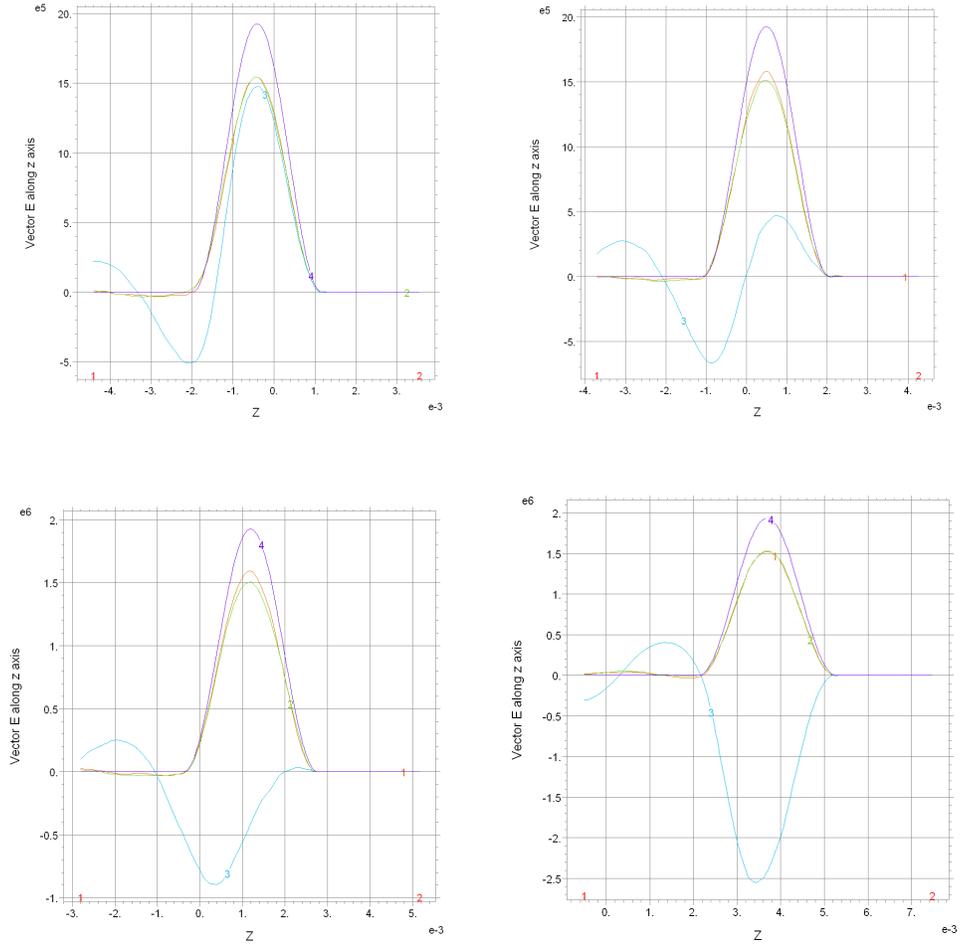

Figure 10. Dynamics of polarity of $E_z$ flip near the half way along *z*. (frames from a movie).

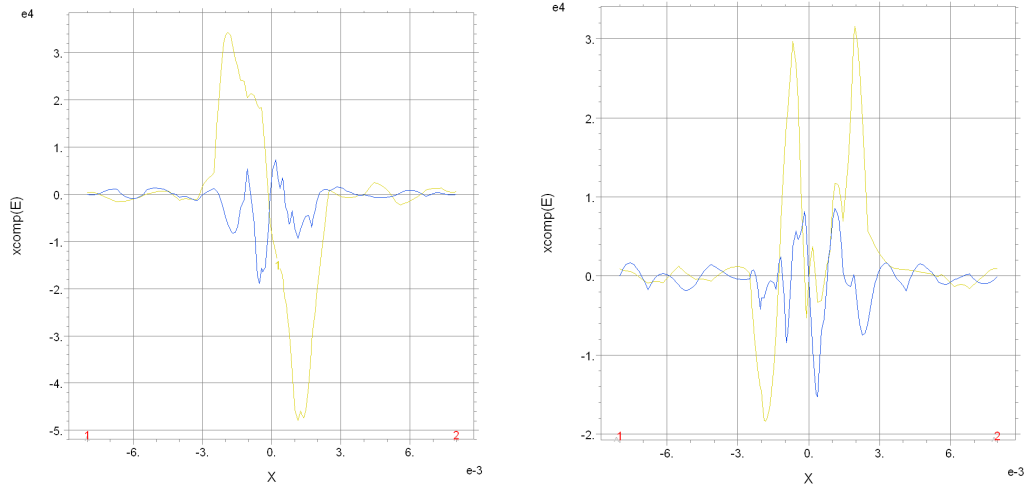

Figure 11. Field across the chamber $-4\sigma_z$ behind the center of a bunch. The central core of chamber has cross section 2.5x2.5 $mm^2$. The level of penetrated field is ~0.001 of maximal field reached at the central cross section, where the field strength is ~6 $MV/m$.

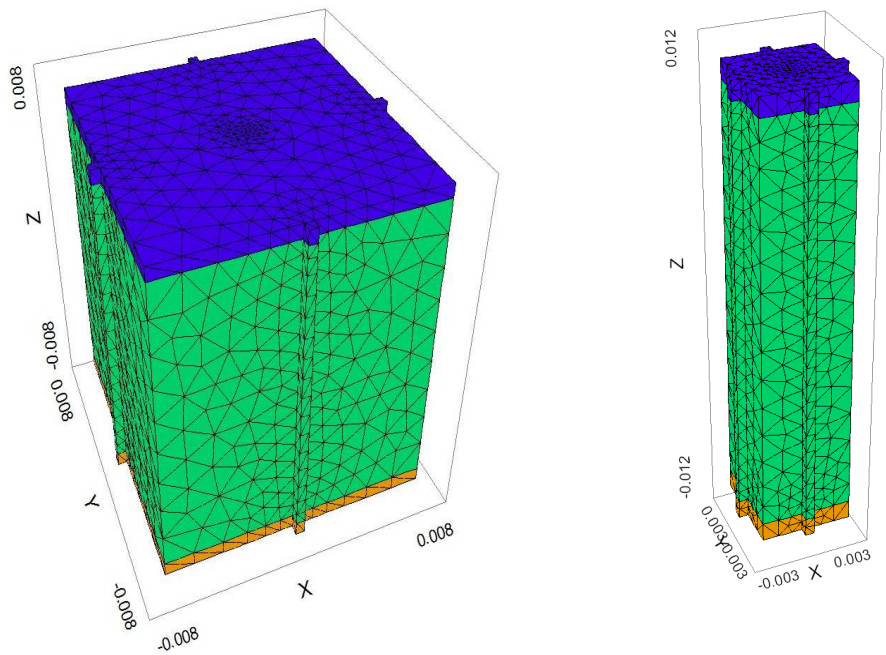

Figure 12. The "Large chamber" and the "Small chamber" from Table 1(see below). The bunch runs through this volume around the central axis $x=0$, $y=0$; dimensions are gives in *meters*.

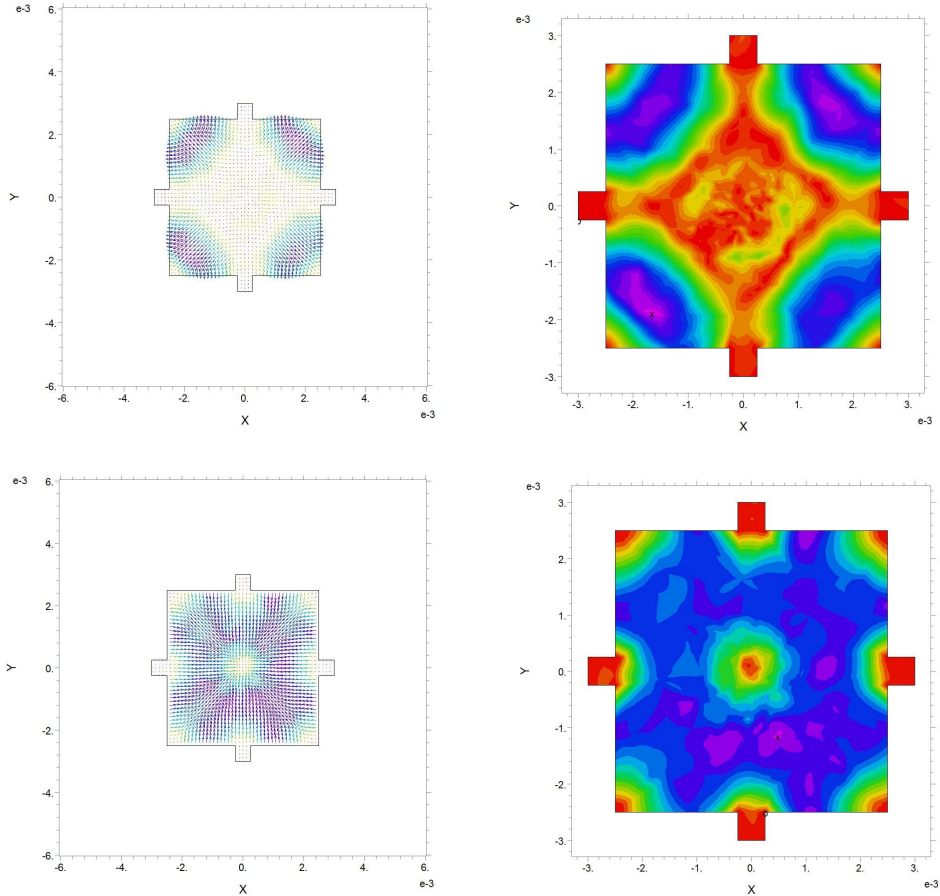

Figure 13. Electric field configurations (vector at the left, field modulus at the right) in a small chamber at $4\sigma_z$ behind the center of bunch (frames from a movie).

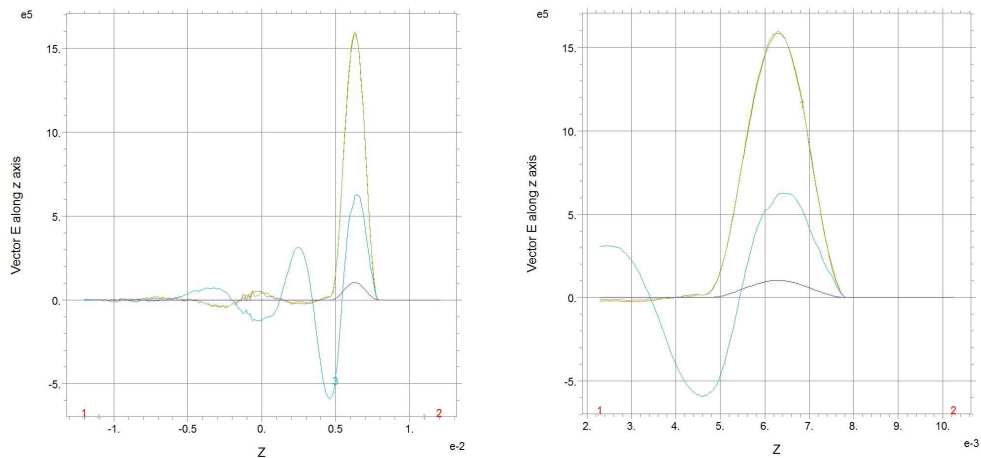

Figure 14. Fields along *z*. At the right the central part of the figure zoomed at $\pm 4\sigma_z$, $\sigma_z = 1mm$. The bunch is moving from the left to the right side (frames from a movie). Field value at maximum reaches 1.5MV/m.

Table 1. Ratio of coherent/incoherent losses for different longitudinal size of bunch;
$K=2$, $\lambda_u = 12mm$, $eN_e=100pQ$, $\sigma_r = 0.3mm$,
the critical angle in formula (10) defined as $\vartheta_m = \vartheta_{coh} \cong \arccos(1-1.4\sigma_b/\lambda_u)$.

| Algorithm\Parameter | E | E | A, E[*)] | E | E | E |
|---|---|---|---|---|---|---|
| $\sigma_z$, mm<br>$\sigma_z/c$ (ps) | 1<br>(3) | 0.5<br>(1.5) | 0.3<br>(1) | 0.1<br>(0.3) | 0.05<br>(0.15) | 0.03<br>(0.1) |
| **Formula (10)** | $6.5 \cdot 10^{-5}$ | $2.57 \cdot 10^{-4}$ | $7.2 \cdot 10^{-4}$ | $6.45 \cdot 10^{-4}$ | 0.0258 | 0.0717 |
| Formula (15) | $3.1 \cdot 10^{-11}$ | $2.6 \cdot 10^{-10}$ | $1.26 \cdot 10^{-9}$ | $3.5 \cdot 10^{-8}$ | $2.8 \cdot 10^{-6}$ | $1.3 \cdot 10^{-6}$ |
| $\Delta\varepsilon_\tau/\Delta\varepsilon_{tot}$ | $1.37 \cdot 10^{-5}$ | $8.0 \cdot 10^{-6}$ | $6.12 \cdot 10^{-5}$ | $5.04 \cdot 10^{-4}$ | $8.3 \cdot 10^{-4}$ | $2 \cdot 10^{-3}$ |
| $\Delta\varepsilon/\Delta\varepsilon_{tot}$ | $3.39 \cdot 10^{-5}$ | $8.6 \cdot 10^{-6}$ | $7.02 \cdot 10^{-4}$ | $3.5 \cdot 10^{-4}$ | $6.1 \cdot 10^{-4}$ | $4.4 \cdot 10^{-3}$ |
| **Large chamber**[**)] | $8.9 \cdot 10^{-5}$ | $2.5 \cdot 10^{-4}$ | $3.4 \cdot 10^{-4}$ | $4.1 \cdot 10^{-3}$ | $1.1 \cdot 10^{-2}$ | 0.065 |
| Small chamber[***)] | $3.0 \cdot 10^{-6}$ | $1.3 \cdot 10^{-5}$ | $1.0 \cdot 10^{-5}$ | $3. \cdot 10^{-5}$ | $1.8 \cdot 10^{-4}$ | $1.2 \cdot 10^{-4}$ |

[*)]Algorithm E referrers to calculation with electric field, algorithm A –to the vector potential.

[**)]Large chamber referrers to $\Delta\varepsilon_\tau/\Delta\varepsilon_{tot}$ calculated for the large vacuum chamber 7.5x7.5 $mm^2$, Fig. 12.

[***)]Small chamber has no extended slits, just central core with 2.5x2.5 $mm^2$ cross section,
 Regular chamber has the "wings" attached to 2.5x2.5 $mm^2$ "Small" chamber; cross section is 8$mm$ wide and 0.5$mm$ thick,
   see Fig.8.

The Table 1 was filled with the following procedure. For some arbitrary moment of time the integrals (25) and (27) ($\Delta\varepsilon = \int dt \int (\vec{j} \cdot \vec{E})dV$ and $\Delta\varepsilon_\tau = \int dt \int (\vec{j} \cdot \vec{E}_\tau)dV$) were calculated and subtracted from the same integrals calculated for the other moment. The same type of calculations for the same moments of time was performed for equation (7) of full non-coherent losses. Then the appropriate ratio of integrals (25) or (27) were divided by the difference of integral (7) calculated for the same moments of time.
   It is clearly seen here that suppression of radiation in a chamber shown in Fig.4, center, is substantial. One can see also, that attenuation of small chamber (chamber without slits) is ~two orders of magnitude stronger, than the "Large" one.
  One comment should be addressed to the formula (10) while comparing with numerical calculations however. In the formula (9) the bunch length substituted in a model of rectangular current distribution. In numerical calculations performed, the sigma of a bunch used defined as (17) or (18). So the effective length defined by these two models might differ 2-3 times, meanwhile the formula (9) is rather sensitive to the exact value of the length. So the real losses calculated with (9) might be up to ±10 times in both directions. Even so, the attenuation of vacuum chamber remains below the factor 10 for the intermediate $\sigma_z$ =0.3$mm$ (1$ps$). As we mentioned in [6] the threshold condition was calculated as $\lambda_{coh} \cong 1.7\sigma_b$ with $\Delta\lambda_{coh} \cong 0.26\lambda_{coh}$

## DISCUSSION AND CONCLUSIONS

For calculation of coherent radiation with formula (25), in some publications the Lineard-Wiechert (L-W) potentials or presentation of electric field such as Jefimenko's form (see [15]) become appointed for this purpose. Problem with this is that these formulas dealing with the point like source (by the Green function for a point-like source). As the integration in (25) is going over the bunch length, where the current exists, one should know the field value at any point inside the bunch. So the field at any point of space $z$ (1D actually) is a superposition of fields from each particle in the bunch. Meanwhile if the reference point (point of observation) is chosen inside the bunch, the divergence $\sim 1/(z-z')$, where $z'$ is a current coordinate inside the bunch, emerges inevitably. To eliminate this divergence in [16] there was proposed just to subtract from the complete expression for electric field the term, responsible for non radiation change of energy. One can see, however, that these terms have different dependence on coordinate ($\sim 1/(z-z')$ for the radiation fields and $\sim 1/(z-z')^2$ for the non radiation fields), so this procedure cannot bring positive (correct) effect for all variety of distances. In addition, the radiation field defined in so called far zone, when the distance to the observation point should be much bigger, than the wavelength of radiation *and* the size of source.

Meanwhile an elegant method for elimination of this divergence was described in [17]. Let we discuss this in a bit more detail here, as this method is pretty universal and powerful and could serve as a basis for further comparison of CSR calculated analytically and numerically with our algorithm in a future. Formula (25) for the power of radiation can be rewritten as

$$P = -\int (\vec{j} \cdot \vec{E}) dV = -\int \vec{j} \cdot (\frac{\partial \vec{A}}{\partial t} + gradU) dV = \int \left( -\vec{j} \cdot \frac{\partial \vec{A}}{\partial t} - \rho \cdot \frac{\partial U}{\partial t} \right) dV + \frac{d}{dt} \int \rho U dV, \quad (32)$$

where the charge conservation condition $div\,\vec{j} + \partial\rho/\partial t = 0$ was used. The last term in (32) could be omitted as it states for the electrostatic change of energy. Meanwhile the solutions of (19) and (21) can be represented as

$$U(\vec{r},t) = \frac{1}{\varepsilon_0} \int \frac{\delta(t'-t+\frac{|\vec{r}-\vec{r}'|}{c})}{|\vec{r}-\vec{r}'|} \rho(\vec{r}',t') dV' dt' \quad (33)$$

$$\vec{A}(\vec{r},t) = \mu_0 \int \frac{\delta(t'-t+\frac{|\vec{r}-\vec{r}'|}{c})}{|\vec{r}-\vec{r}'|} \vec{j}(\vec{r}',t') dV' dt' \quad (34)$$

Now, by introduction of

$$\delta(t) = \frac{1}{2\pi} \int e^{i\omega t} d\omega, \quad (35)$$

one can transform (33) and (34) into

$$U(\vec{r},t) = \frac{1}{2\pi\varepsilon_0} \int \frac{e^{i\omega(t'-t)+i\omega|\vec{r}-\vec{r}'|/c}}{|\vec{r}-\vec{r}'|} \rho(\vec{r}',t') dV' dt' d\omega \quad (36)$$

$$\vec{A}(\vec{r},t) = \frac{\mu_0}{2\pi} \int \frac{e^{i\omega(t'-t)+i\omega|\vec{r}-\vec{r}'|/c}}{|\vec{r}-\vec{r}'|} \vec{j}(\vec{r}',t') dV' dt' d\omega. \quad (37)$$

Next, by using the equity

$$\frac{\sin(\omega|\vec{r}-\vec{r}'|/c)}{|\vec{r}-\vec{r}'|} = \frac{\omega}{c}\int e^{i\omega\cdot\vec{n}(\vec{r}-\vec{r}')/c}\frac{do}{4\pi} , \qquad (38)$$

one can rewrite (36) and (37) as the following [17]

$$U(\vec{r},t) = \frac{i}{2\pi\varepsilon_0}\int e^{i\omega(t'-t)+i\omega\vec{n}(\vec{r}-\vec{r}')/c}\rho(\vec{r}',t')dV'dt'\omega d\omega\frac{do}{4\pi} \qquad (39)$$

$$\vec{A}(\vec{r},t) = \frac{i\mu_0}{2\pi}\int e^{i\omega(t'-t)+i\omega\vec{n}(\vec{r}-\vec{r}')/c}\vec{j}(\vec{r}',t')dV'dt'\omega d\omega\frac{do}{4\pi} . \qquad (40)$$

By substitution (39) (40) into (32) one can get the total radiated power as

$$P(t) = \frac{1}{2\pi c}\int e^{i\omega[(t'-t)+\vec{n}(\vec{r}-\vec{r}')/c]}\left[\rho(\vec{r},t)\rho(\vec{r}',t') - \frac{1}{c^2}\vec{j}(\vec{r},t)\vec{j}(\vec{r}',t')\right]dV'dVdt'\omega^2 d\omega\frac{do}{4\pi} = \int P(\vec{n},t)do \qquad (41)$$

where $P(\vec{n},t)$ represents the power radiated per unit solid angle around direction $\vec{n}$ at the time $t$. For the bunch density $\rho(\vec{r},t)$ and for the current density $\vec{j}(\vec{r},t)$ one can use the formulas (16), (17), (19). Formula (41) can be rewritten also as

$$P(t) = \int d\omega \int P(\vec{n},\omega,t)do , \qquad (42)$$

where $P(\vec{n},\omega,t)$ stands for the power radiated per unit solid angle around direction $\vec{n}$ at the time $t$ in a frequency interval $d\omega$ around frequency $\omega$. Substitute in (41) expressions for the charge density and current (16), (29) one can rewrite it as

$$P(t) = \frac{1}{2\pi c}\int e^{i\omega[(t'-t)+\frac{\vec{n}(\vec{r}-\vec{r}')}{c}]}\rho(\vec{r}_\perp,z-v_z t)\rho(\vec{r}'_\perp,z'-v_z t')\cdot\left[1-\frac{1}{c^2}\vec{v}(\vec{r},t)\vec{v}(\vec{r}',t')\right]dV'dVdt'\omega^2 d\omega\frac{do}{4\pi} \qquad (43)$$

One can see that this approach does not require any "regularization" as it gives direct value for the radiated power. Formula (43) valid for the bunch moving in a *free space*, as the boundary conditions are not taken into account (which affects (33) and (34) basically, where instead of $\delta(t'-t+\frac{|\vec{r}-\vec{r}'|}{c})/|\vec{r}-\vec{r}'|$ an appropriate Green function should be used).

The approach, accepted in our current publication, deals with numerical solution of (19), (21) or (28) together with (22). As the driving term (the source) has no singularities, the solution does not suffer from the divergence, associated with infinitely small distance between the observation point and the source point. For the purposes of power calculation, the algorithm for calculation of vector and scalar potential (19), (21) working as well as the algorithm with direct calculation of electric field (28). Algorithm with direct calculation of electric field exposed in this current publication working ~10 times faster than the one associated with the vector potential.

Interesting dynamical phenomena discovered while the numerical calculations performed. Namely, establishment of electric field, while the bunch enters the vacuum chamber is going with reflection from the walls, so the sign of electric field across the chamber behind the bunch periodically changes its polarity. Moreover, the periodic energy exchange between the bunch and the coherent field is going all the time, while the bunch is passing the region with undulator field. This is a manifestation of the fact, that although the formation length is equal to the undulator period, only after passage of ~137 periods the real photon becomes radiated. So the energy exchange between the bunch and the field can be considered as radiation of virtual photons at this distance.

There is strong indication, that the coherent phenomena in an undulator are sensitive to the *transverse* beam size, rather than the longitudinal one, at least for some dimensions ratio. The physical sense is clear: in a moving system of coordinates, the bunch length together with the wavelength of undulator field becomes shrunk by a factor $1/\gamma$, so the *transverse* size defines the coherence process in this (moving) system now, as the bunch length remains much smaller, than the wavelength of oscillation in a moving frame. For a typical case, the radius of trajectory (see Fig. A1) is $r = \lambdabar_u K / \gamma \cong 5 \cdot 10^{-5} cm$ which is much smaller, than the characteristic transverse beam size $\sigma_b \cong \sqrt{\gamma \varepsilon \cdot \beta / \gamma} \cong 3 \cdot 10^{-3} cm$, where $\gamma\varepsilon \cong 10^{-6} m \cdot rad$ is an invariant emittance, $\beta \cong 10m$ stands for the envelop function. The wavelength of radiation in a moving frame is $\lambda' \cong \lambda_u / \gamma = c 2\pi r / V$. So that is why the exact value of ratio $r/\sigma_b$ is important here. This basically means that in general, that the coherent phenomenon in an undulator is a 3D problem.

Coherent radiation reduction is ~20 in a chamber with wide slits (as in delta-undulator [9]). In addition, the CSR is sensitive to the exact dimensions of these slits, as they demonstrate resonant properties. For a bunch with charge 100*pC*, the time duration of 100*fs*, the ratio of energy lost by coherent radiation to the energy lost incoherently may reach approximately 2.2 for the undulator with $\lambda_u = 2.4 cm$, *K*=2. For the central core with dimensions which are less, than the critical ones, the radiation propagates exclusively in the slits. In this case the *transverse* size of the beam becomes dominating parameter. This is in agreement with the physical expectations, as in this case the radiation is emitted to the sides ($\theta \approx \pi/2$), so that transverse dimensions define the coherent threshold.

General conclusion is that CSR is tolerable in all rage of parameters suggested for ERL.

*Appendix*
*Derivation of formula for angular distribution of radiation for arbitrary angles*

Although the formula (3) was derived a long time ago [2]-[5] we represent here its derivation again for completeness, as some original publications is difficult to find now, and in mostly tutorials it is derived for small angles only, when

$$\frac{1}{1-\beta_\| Cos\vartheta} \to \frac{1}{\frac{1}{2\gamma^2}(1+\gamma^2\vartheta^2+K^2)}.$$

As in the formula for intensity this factor appears in a third power, the discrepancy could be significant.

The formula of our interest (3) will be derived in the following two-step way. At the first step it will be derived for a system, which is moving with average velocity $\bar{v} = c\bar{\beta} \cong c\beta_\|$. In this system the electron performs simple circular motion (for a helical undulator/wiggler), so the formula for radiation is the one derived first by Schott [12]. At the second step we shall transform the angular distribution into the Lab frame with the help of formulas for relativistic transformation of angles. Let us follow this plan.

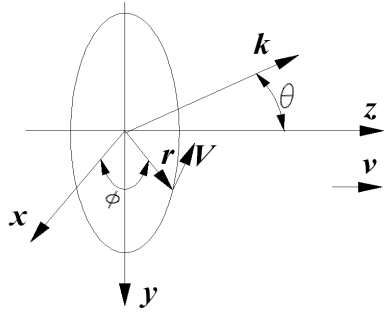

Figure A1. The system of coordinates, which is moving with average velocity $\bar{v} = c\bar{\beta} \cong c\beta_\parallel$. In a moving frame, the particle is spinning at radius *r* with Velocity *V*.

Equation for the vector potential (19) can be considered in a moving system of coordinates, Fig.A1. As the motion of particle in a moving frame is periodic, the vector potential in this moving frame has a form [7]

$$\vec{A}_n = \frac{e}{cT} \int_{-\infty}^{\infty} \frac{\vec{V}(t)}{R(t)} e^{in\omega_0[t+R(t)/c]} dt = \frac{e}{cT} \int_{-\infty}^{\infty} e^{in\omega_0[t+R(t)/c]t} d\vec{r}(t) \quad (A1)$$

where $R(t)$ is an instant distance from the particle to the observation point, $\vec{V}(t) = \dot{\vec{r}}(t)$ is an instant velocity of particle spinning in a moving frame, $T = 2\pi/\omega_0 = 2\pi r/V$ and n=1.2.3,… numerates the harmonics of revolution frequency. The last equation can be expanded further while representing $R(t) = R_0 - \vec{r}\vec{n}$ (here $\vec{n}$ is an unit vector in direction to observation point) and taking into account that coordinates of particle are $x = r\cos\omega_0 t$, $y = r\sin\omega_0 t$, $z = 0^1$ one gets

$$\vec{A}_n = \frac{e^{ikR_0}}{cR_0 T} \int_{-\infty}^{\infty} e^{i[n\omega_0 t - \vec{k}\vec{r}]} d\vec{r}(t) , \quad (A2)$$

where $k = n\omega_0/c = nV/cr$. As $\vec{k}\vec{r} = kr\sin\theta\sin\phi = \frac{nV}{c}\sin\theta\sin\phi$, $\phi = \omega_0 t = (2\pi V/r)\cdot t$, then

$$A_{xn} = -\frac{eV\, e^{ikR_0}}{2\pi c R_0} \int_0^\pi e^{in[\phi - \frac{V}{c}\sin\theta\sin\phi]} \sin\phi\, d\phi \quad (A3)$$

$$A_{yn} = \frac{eV\, e^{ikR_0}}{2\pi c R_0} \int_0^\pi e^{in[\phi - \frac{V}{c}\sin\theta\sin\phi]} \cos\phi\, d\phi \quad (A4)$$

Integrals (A3) (A4) can be expressed through the Bessel function and its derivative [13]

---

[1] For the helical motion of our interest $z = \beta_\parallel ct$ so in the following formula (A2) for the vector potential a longitudinal component appears if calculated in the Lab frame; $dz = \beta_\parallel c dt = \beta_\parallel c d\phi/\omega_0$. This makes calculation a bit more complicated that is why we prefer the way proclaimed (relativistic transformation of Schott's formula).

$$A_{xn} = -\frac{ieV\,e^{ikR_0}}{cR_0} J'_n\!\left(\frac{nV}{c}\sin\theta\right), \qquad A_{yn} = \frac{eV\,e^{ikR_0}}{cR_0 \sin\theta} J_n\!\left(\frac{nV}{c}\sin\theta\right) \tag{A5}$$

As the intensity of radiation defined by [7]

$$\frac{dI'_n}{do'} = \frac{cR_0^2}{2\pi}\left|\vec{k}\times\vec{A}_n\right|^2 = \frac{cR_0^2}{2\pi} k^2 \left[A_{xn}^2 + A_{yn}^2 \cos^2\theta\right], \tag{A6}$$

where the accent $I', do'$ means, that the intensity and spherical angle calculated in a moving frame. Substitute here expressions for the components of vector potential (A3) and (A4) we arrive at Schott's formula[2] (for the right-hand circular polarization)

$$\frac{dI'_n}{do'} = \frac{e^2 \omega^2 n^2}{c}\left[\cot^2\theta \cdot J_n^2\!\left(\frac{nV}{c}\sin\theta\right) + \frac{V^2}{c^2} J'^2_n\!\left(\frac{nV}{c}\sin\theta\right)\right] \tag{A7}$$

Now let us transform this formula into the Lab frame. According to the rules for transformation of angles from the moving system ($\theta, \phi$) to the Lab one ($\vartheta, \varphi$) [7]

$$\sin\theta = \frac{\sqrt{1-\beta_\parallel^2}}{1-\beta_\parallel \cos\vartheta}\sin\vartheta, \qquad \cos\theta = \frac{\cos\vartheta - \beta_\parallel}{1-\beta_\parallel \cos\vartheta}, \qquad \phi = \varphi \tag{A8}$$

Intensity defined as the energy $\mathcal{E}$ lost per second

$$\frac{dI'}{do'} = \frac{d^2\mathcal{E}'}{do'dt'} = f(\sin\theta,\phi), \quad do' = d(\cos\theta)d\phi = \frac{1-\beta_\parallel^2}{\left(1-\beta_\parallel \cos\vartheta\right)^2} do, \tag{A9}$$

where the function $f$ defined accordingly to (A7). The energy and time transforming as

$$d\mathcal{E}' = d\mathcal{E}\cdot\frac{1-\beta_\parallel \cos\vartheta}{\sqrt{1-\beta_\parallel^2}}, \qquad dt' = dt\cdot\sqrt{1-\beta_\parallel^2}, \tag{A10}$$

so the angular distribution in Lab frame comes to [7]

$$\frac{dI}{do} = \frac{d^2\mathcal{E}}{dodt} = \frac{\left(1-\beta_\parallel^2\right)^2}{\left(1-\beta_\parallel^2 \cos\vartheta\right)^3} f\!\left(\frac{\sqrt{1-\beta_\parallel^2}\cdot\sin\vartheta}{1-\beta_\parallel \cos\vartheta},\varphi\right). \tag{A11}$$

One can see from (A10) that the energy radiated per unit time in a particular direction counted in the moving and in its own system of reference is not an invariant under Lorentz transformation, as according to (A10)

$$\frac{d\mathcal{E}'}{dt'} = \frac{d\mathcal{E}}{dt}\cdot\frac{1-\beta_\parallel \cos\vartheta}{1-\beta_\parallel^2} \xrightarrow{\vartheta=0} = \frac{d\mathcal{E}}{dt}\cdot\frac{1-\beta_\parallel}{1-\beta_\parallel^2} \cong \frac{1}{2}\frac{d\mathcal{E}}{dt}$$

---

[2] We would like to attract attention, that this formula differs from the formula (74.8) from [7] due to the difference in definition of $\theta$ by $\pi/2$.

Taking into account, that $\frac{V}{c} = \beta_\perp / \sqrt{1-\beta_\parallel^2}$ and expanding $f$ according to (A7) we arrive at

$$\frac{dI_n}{do} = \frac{e^2 n^2 \omega_0^2}{c(1-\beta_\parallel \cos\vartheta)^3}\left[\beta_\perp^2 J_n'^2\left(\frac{n\beta_\perp Sin\vartheta}{1-\beta_\parallel Cos\vartheta}\right) + \frac{(Cos\vartheta - \beta_\parallel)^2}{Sin^2\vartheta} J_n^2\left(\frac{n\beta_\perp Sin\vartheta}{1-\beta_\parallel Cos\vartheta}\right)\right], \quad (A12)$$

which could be transformed into (3). One can see easily, that if $\beta_\parallel = 0$, then (A12) coincides with Schott's formula (A7).